\documentclass[aps,prb,twocolumn]{revtex4-2}
\usepackage{graphicx,bm}

\begin{document}

\title{Magnetization dynamics with time-dependent spin-density functional theory: significance of exchange-correlation torques}

\author{Daniel Hill}
\author{Justin Shotton}
\author{Carsten A. Ullrich}
\email{ullrichc@missouri.edu}
\affiliation{Department of Physics and Astronomy, University of Missouri, Columbia, Missouri 65211, USA}

\date{\today}

\begin{abstract}
In spin-density-functional theory (SDFT) for noncollinear magnetic materials, the Kohn-Sham system features exchange-correlation (xc) scalar
potentials and magnetic fields. The significance of the xc magnetic fields is not very well explored; in particular,
they can give rise to local torques on the magnetization, which are absent in standard local and semilocal approximations.
Exact benchmark solutions for a five-site extended Hubbard lattice at half filling and in the presence of spin-orbit coupling are compared with SDFT results
obtained using orbital-dependent exchange-only approximations. The magnetization dynamics following short-pulse excitations is found to be reasonably
well described in the exchange-only approximation for weak to moderate interactions. For stronger interactions and near transitions between magnetically
ordered and frustrated phases, exchange and correlation torques tend to compensate each other and must both be accounted for.
\end{abstract}

\maketitle

\newcommand{\bfr}{{\bf r}}
\newcommand{\bfm}{{\bf m}}
\newcommand{\bfB}{{\bf B}}
\newcommand{\bfb}{{\bf b}}
\newcommand{\bfJ}{{\bf J}}
\newcommand{\tens}[1]{\underline{\underline{#1}}}

\section{\label{sec:intro1}Introduction}

Spin dynamics in magnetic systems is a research area of much current activity. Spintronics \cite{Zutic2004}, which is concerned with
the manipulation of electronic spins, spin currents, spin textures, and spin excitations, has created a wealth of scientific
knowledge and many avenues for new technologies. Prominent examples are spin waves for encoding and transmitting information
(magnonics) \cite{Rezende2020,Barman2021}, skyrmions for magnetic information storage \cite{Nagaosa2013,Fert2017,Zhang2017,Gobel2021,Chen2022}, and single-spin qubits for quantum computation \cite{Vandersypen2017}. Another related area of much interest is ultrafast demagnetization induced by femtosecond laser pulses
\cite{Bovensiepen2009,Krieger2015,GPZhang2016,GPZhang2016a,Acharya2020}.

Computational approaches to simulate magnetization dynamics in a wide variety of systems are typically based
on the Landau-Lifshitz-Gilbert (LLG) equation of motion \cite{Lifshitz,Eriksson2017}. The LLG equation provides a classical
description of the time evolution of the magnetization vector $\bfm(t)$ in response to a time-dependent perturbation 
(typically, a short pulse or a periodic driving field)
or evolving from a nonequilibrium initial state. Materials properties such as anisotropy, deformations, strain,
and various forms of damping can be built into the LLG approach via phenomenological or ``second-principles'' parameters.

In this paper, we are less concerned with these specific materials properties; instead of LLG we will use a fully quantum mechanical
description of the electronic charge and spin degrees of freedom, and our focus will be specifically on the impact of
electron-electron interactions on the magnetization dynamics. To be more clear, we consider
a system of $N$ interacting electrons under the influence of a time-dependent scalar potential $V(\bfr,t)$ and
a time-dependent magnetic field $\bfB(\bfr,t)$ which couples only to the electron spin (and not to orbital motion).
The associated many-body Hamiltonian is given by
\begin{eqnarray} \label{H}
\hat H &=& \sum_{j}^N\left[-\frac{\nabla_j^2}{2} + V(\bfr_j,t) + \bm{\sigma}_j\cdot \bfB(\bfr_j,t)  \right]
\nonumber\\
&+&
\frac{1}{2}\sum_{j\ne k}^N \frac{1}{|\bfr_j - \bfr_k|} \:,
\end{eqnarray}
where $\bm{\sigma}_j$ is the vector of Pauli matrices acting on the spin of the $j$th electron,
and we define the magnetic field strength such that the Bohr magneton, $\mu_B = e \hbar/2m$, does not explicitly appear in the Hamiltonian
$\hat H$. We use atomic units ($e = m = \hbar = 4\pi \epsilon_0 = 1$) throughout.

From the Heisenberg equation of motion for $\hat H$, Capelle {\em et al.} showed that the magnetization has the following time evolution \cite{Capelle2001}:
\begin{equation} \label{m}
\frac{d \bfm(\bfr,t)}{dt}  + \hat \nabla \cdot \bfJ(\bfr,t) = \bfm(\bfr,t) \times \bfB(\bfr,t) \:,
\end{equation}
where $\bfJ(\bfr,t)$ is the spin-current tensor. Equation (\ref{m}) is exact but not very helpful in practice since $\bfJ(\bfr,t)$ requires
the many-body wave function associated with $\hat H$.
A more practical (but still in principle exact) alternative is time-dependent spin-density functional theory (TD-SDFT).
The idea of TD-SDFT is to consider an auxiliary system of noninteracting fermions, acted upon by an ``effective'' scalar potential and
magnetic field, $V_{\rm eff}(\bfr,t)$ and $\bfB_{\rm eff}(\bfr,t)$, such that the same density $n(\bfr,t)$ and magnetization
$\bfm(\bfr,t)$ are produced as in the physical system. The resulting equation of motion, the TD-SDFT counterpart to Eq. (\ref{m}),
is \cite{Capelle2001}
\begin{equation} \label{mks}
\frac{d \bfm(\bfr,t)}{dt}  + \hat \nabla \cdot \bfJ_{\rm KS}(\bfr,t) = \bfm(\bfr,t) \times \bfB_{\rm eff}(\bfr,t) \:.
\end{equation}
Here, $\bfJ_{\rm KS}(\bfr,t)$ is the Kohn-Sham spin-current tensor, which is easily determined from the noninteracting wave function,
and the effective magnetic field is defined as $\bfB_{\rm eff}(\bfr,t) = \bfB(\bfr,t) + \bfB_{\rm xc}(\bfr,t)$, where
the exchange-correlation (xc) magnetic field $\bfB_{\rm xc}$ is a functional of the density and magnetization.
Formally, $\bfm(\bfr,t)$ is the same in Eqs. (\ref{m}) and (\ref{mks}),  but $\bfJ$ and $\bfJ_{\rm KS}$ are in general different
(the difference lies in the transverse component). Thus, the so-called xc torque,
\begin{equation} \label{tau}
\bm{\tau}_{\rm xc}(\bfr,t) = \bfm(\bfr,t) \times \bfB_{\rm xc}(\bfr,t) \:,
\end{equation}
ensures that TD-SDFT produces the correct magnetization dynamics \cite{Capelle2001}.

While all of this is clear at the formal level, the exact form of $\bfB_{\rm xc}$ is unknown and must be approximated in practice.
This immediately raises several questions: which approximations of $\bfB_{\rm xc}$ are available, and do they produce xc torques?
And, how important are the xc torques for the magnetization dynamics?

A number of approximations for $\bfB_{\rm xc}$ have been derived within ground-state SDFT for noncollinear magnetism \cite{Barth1972,Gunnarsson1976,Gidopoulos2007}; via the adiabatic approximation, they immediately carry over to TD-SDFT.
The most widely used approach, pioneered by K\"ubler {\em et al.} \cite{Kubler1988,Sandratskii1998} and implemented in many popular
electronic structure codes, is to use standard local or semilocal xc functionals such as the local spin-density approximation (LSDA) or
generalized gradient approximations (GGAs), and assume a local spin quantization axis which is aligned with the
local magnetization vector $\bfm(\bfr,t)$; this produces a $\bfB_{\rm xc}(\bfr,t)$ that is parallel to $\bfm(\bfr,t)$ everywhere.
We see right away from Eq. (\ref{tau}) that this class of approximations does not produce any xc torques.

Approximations for $\bfB_{\rm xc}$ that do include xc torque effects can be constructed in several ways. Existing local and semilocal functionals
(LSDA and GGAs) have been modified \cite{Katsnelson2003,Peralta2007,Scalmani2012,Bulik2013} or used in a source-free construction \cite{Sharma2018}, and new gradient-corrected functionals were constructed based using the spin-spiral state of the electron gas as reference system \cite{Kleinman1999,Eich2013a,Eich2013b}.
More consistent derivations of xc meta-GGAs, starting from noncollinear generalizations of the exchange hole
and the two-body density matrix, were recently presented \cite{Pittalis2017,Tancogne2022}. Various orbital-dependent functionals were generalized to the case of
noncollinear magnetization \cite{Sharma2007,Capelle2010,Ullrich2018}.

Existing applications of ground-state SDFT to noncollinear magnetic materials \cite{Sharma2007,Scalmani2012,Bulik2013} and model
systems \cite{Pluhar2019} seem to suggest that xc torques are of relatively minor importance for magnetic structure and energetics,
although the torques themselves may not be insignificant \cite{Tancogne2022}.
On the other hand, there are good reasons to expect that xc torques will be more impactful for magnetization dynamics: they
explicitly appear in the equation of motion, Eq. (\ref{mks}), and even if $\bm{\tau}_{\rm xc}(\bfr,t)$ is relatively small at a given
$\bfr$ and $t$, its effect can accumulate over time.
So far, however, there has been no systematic attempt to assess this hypothesis. We are only aware of one study in the literature, where
Dewhurst {\em et al.} \cite{Dewhurst2018} used their source-free $\bfB_{\rm xc}$ functional to simulate laser-induced spin dynamics
in bulk Co and Ni and Co-Pt and Ni-Pt interfaces. They found that xc torques were significant only if they are not overshadowed
by magnetic anisotropy effects (i.e., in bulk, and not at interfaces), and that they give rise to rather slow spin rotation compared
to other forms of spin dynamics, induced optically or via spin-orbit coupling (SOC).

In this paper, our goal is to assess the importance of xc torques in frustrated magnetic systems.
Exchange-frustrated solids such as spin glasses and kagome antiferromagnetic lattices
are characterized by many competing noncollinear spin configurations and quantum spin liquid phases \cite{Balents2010,Zhou2017,Broholm2020},
and may therefore exhibit an enhanced sensitivity to subtle
xc torque effects. Needless to say, extended spin frustrated solids are challenging to describe, and exact or quasi-exact benchmark results
are hard to come by. We will therefore limit ourselves to small model systems which capture the spirit of spin frustration and yet are
computationally manageable.

Here, we will consider small Hubbard-type model systems along similar lines as in our earlier studies \cite{Ullrich2018,Pluhar2019,Ullrich2019};
by including SOC we can generate intrinsically noncollinear ground states.
In particular, we will focus on a five-site half-filled Hubbard bowtie as a minimal model for studying
xc torque effects in the presence of magnetic frustration.  We will generate both  exact and SDFT
phase diagrams of spin configurations for this system and explore the spin dynamics for different configurations in the phase
diagram. The TD-SDFT treatment will be based on orbital-dependent exchange-only functionals, and we will compare with exact solutions of the
many-body time-dependent Schr\"odinger equation. Focusing on a few representative case studies, we will gain insight into the
significance of xc torques in different regimes.

The paper is organized as follows. In Sec.~\ref{sec:model2} the extended Hubbard model and the SDFT framework are introduced
and the exact and SDFT magnetic phase diagrams are discussed. In Sec.~\ref{sec:III} we describe some technical aspects of the TD-SDFT modeling
such as the choice of initial state.
In Sec.~\ref{sec:res4} the results of exact diagonalization and SDFT models are compared for the cases with moderate to strong correlations and non-local interactions.
Conclusions are given in Sec.~\ref{sec:con5}.

\begin{figure}
\includegraphics[width=\columnwidth]{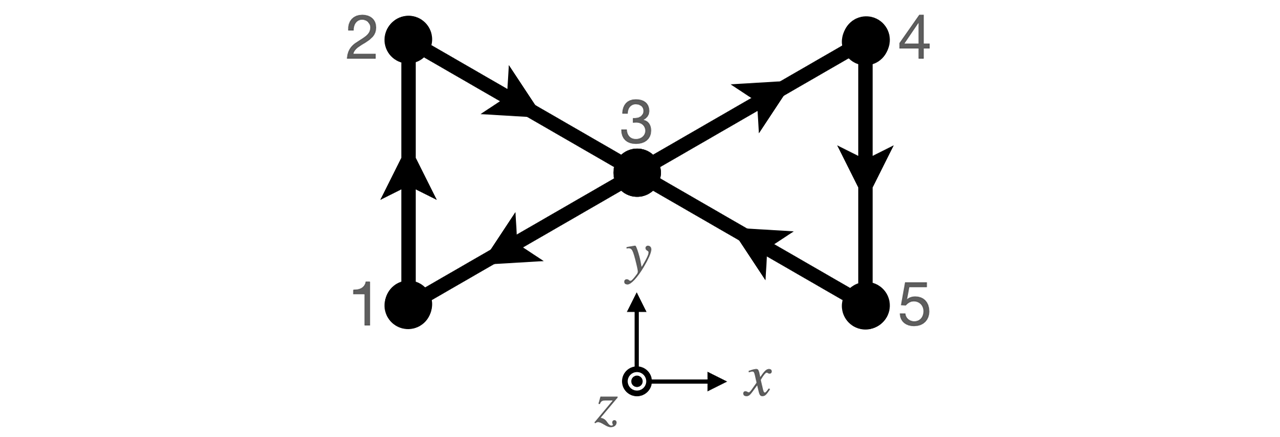}
\caption{\label{fig:geo} Geometry of the 5-site Hubbard cluster used in this work. The arrows indicate the ordering of the nearest-neighbor sum in Eq. (\ref{eqn:hop}),
accounting for the directional hopping due to SOC.}
\end{figure}

\section{\label{sec:model2} Exact and SDFT magnetic structure of Hubbard clusters}

\subsection{Definition of the model}

In this paper we limit ourselves to (TD-)SDFT in the exchange-only approximation. 
As discussed earlier \cite{Pluhar2019}, the standard Hubbard model with on-site
interactions does not give rise to any exchange torques. If one wishes to study exchange torque effects it is necessary to work with an extended Hubbard model instead.
We will consider, in the following, a half-filled 5-site Hubbard cluster in a bowtie shape, as shown in Fig. \ref{fig:geo}.
Here, we go beyond Ref. \cite{Pluhar2019} and include SOC through a modification of the kinetic-energy operator, where the hopping term
becomes complex and the hopping acquires a directionality \cite{Kaplan1983,Tabrizi2019}.
Thus, our inhomogeneous extended Hubbard model with SOC is described by the Hamiltonian
\begin{equation}
\label{eqn:ham}
    \hat H_{\rm model} = \hat H_T + \hat H_U + \hat H_{\rm ext} \:.
\end{equation}
The first term is a hopping term with SOC absorbed into a spin dependent phase factor,
\begin{equation}
\label{eqn:hop}
    \hat H_T = -t_h \sum_{\langle j,j'\rangle} \sum_{\sigma} e^{-i\sigma \theta} c^{\dag}_{j\sigma} c_{j'\sigma} + h.c. ,
\end{equation}
where $h.c.$ stands for Hermitian conjugate. Here, $t_h = \sqrt{T^2+C^2}$ is the generalized hopping strength parameter which depends on nearest neighbor hopping strength $T$ and spin orbit coupling $C$, $j$ is the site index for the geometry shown in Fig.~\ref{fig:geo},
$c_{j\sigma}$ is the annihilation operator for an electron of spin $\sigma$ at site $j$, the brackets $\langle \dots \rangle $ denote an ordered sum over nearest neighbors
with the order indicated by the arrows in Fig.~\ref{fig:geo}, and $\sigma =\pm 1$ labels spin-up and -down. Furthermore, $\theta$ is the SOC angle which parameterizes
the strength of the SOC parameter $C$ relative to the conventional hopping term $T$  \cite{Pixley2016,Li2020,Hill2021}.

The second term in the model Hamiltonian (\ref{eqn:ham}) comprises the on-site and nearest-neighbor interaction terms,
\begin{equation}
    \hat H_U = U_0 \sum_j n_{j\uparrow}  n_{j\downarrow} + U_1 \sum_{\langle j,j'\rangle} \sum_{\sigma, \sigma'}  n_{j\sigma}  n_{j'\sigma'} \:,
\end{equation}
where $n_{j \sigma} = c^{\dag}_{j\sigma} c_{j\sigma}$ is the spin $\sigma$ particle number density at site $j$,
and $U_0$ and $U_1$ are the on-site and nearest-neighbor repulsion strengths, respectively.
For the purposes of this paper, we set $U_1=\frac{1}{2}U_0$, a fairly typical choice for modeling real materials \cite{Strack1993}, and we restrict the hopping parameter $t_h$ and on-site interaction parameter $U_0$ to be of similar orders of magnitude. Finite nonlocal interactions are necessary for nontrivial exchange torques, but we avoid the much stronger interactions regime because the charge degrees of freedom tend to freeze out as $U_0$ and $U_1$ become large, resulting in the dynamics being dominated by a simpler pure-spin low-energy effective model.

Lastly, $\hat H_{\rm ext}$ contains the couplings to the external potential and external magnetic field,
\begin{equation}
   \hat  H_{\rm ext} = \sum_{j} ( V_j n_j +  \bfB_j \cdot \bfm_j) \:,
\end{equation}
where $V_j$ is the scalar potential and $\bfB_j$ is the magnetic field on site $j$, the total density is $n_j = n_{j\uparrow} + n_{j\downarrow}$,
and the magnetization is given by
$\bfm_j = \sum_{\sigma,\sigma'} c^{\dag}_{j\sigma} \vec{\sigma}_{\sigma \sigma'} c_{j\sigma'} $ with $\vec{\sigma} = (\sigma_x,\sigma_y,\sigma_z)$ denoting a vector composed of the Pauli matrices.
We keep the external field parameters each less than the on-site interaction and hopping, $ V_j ,|\bfB_j| < U_0,t_h$. These external field parameters are not strictly set to zero because they can be used to break degeneracy in order to fix a symmetry breaking state, and because, as discussed in Section \ref{sec:III}, small variation of these parameters in the exact model is found to be useful in matching the SDFT initial state and the exact initial state more accurately.

\begin{figure*}
\includegraphics[width=\textwidth]{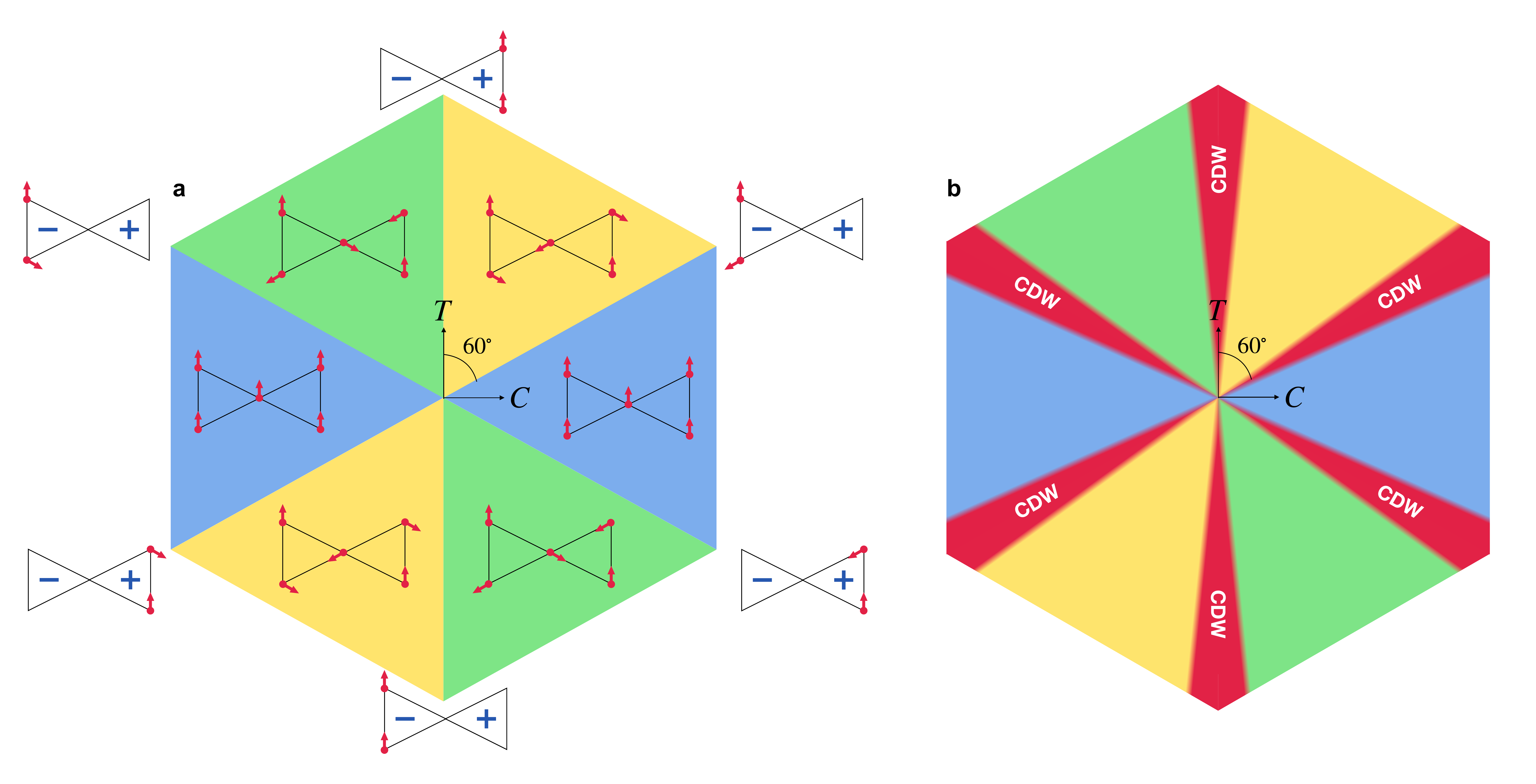}
\caption{\label{fig:PD} (a) Magnetic phase diagram of the half-filled 5-site Hubbard model, obtained using exact diagonalization. The red arrows indicate the relative in-plane spin direction of the state depicted (taken at the midpoint angle between the phase boundaries). The blue pluses and minuses indicate the direction of electric polarization for the CDW critical angle states for the specific spin arrangement shown.
(b) Corresponding magnetic phase diagram using exchange-only SDFT, showing the broadening of the phase boundary states. The phase diagram has approximately the same states as for the exact diagonalization phase diagram, but the critical angles, where a CDW occurs, acquire a width of a few degrees.}
\end{figure*}

\subsection{Magnetic phase diagram of the Hubbard bowtie}

We use exact diagonalization of $\hat H_{\rm model}$ to construct benchmark solutions with which to compare our SDFT results. Figure \ref{fig:PD}a shows the exact phase diagram
of the half-filled Hubbard bowtie in a plane whose $x-y$ axes are defined by $C = (t_h/U_0) \sin\theta$ and $T=(t_h/U_0) \cos\theta$; the SOC angle $\theta$ is here
measured with respect to the kinetic energy axis. Similar phase diagrams
for the half-filled Hubbard trimer were obtained by Tabrizi {\em et al.} \cite{Tabrizi2019}.
Within the above specified regime the model has a phase transition at $\theta_c = n \pi/3$ for any integer $n$. For the case of zero external fields, the ground state of the 5-site model at half filling is degenerate and magnetically ordered with a nontrivial noncollinear spin structure (except at isolated points in the phase diagram where the spins are ferromagnetically aligned) indicating magnetic frustration.

On the phase boundary, $\theta_c$, the ground state exhibits a symmetry breaking charge density wave (CDW) in the form of a spontaneous charge polarization along the $x$-axis of Fig.~\ref{fig:geo}.
In Fig. \ref{fig:PD}a the states shown outside the phase diagram image are the states at the critical angles $\theta_c$. A specific choice of charge polarization is depicted in order to show the corresponding spin state. The sites with no spin indicated do not necessarily have zero magnetic moment, but it tends to be orders of magnitude smaller.
The states shown inside the shaded segments of the phase diagram are those of the midpoint angles between the phase boundaries, e.g. $\theta = 30^{\circ}$. As $\theta$ changes, the relative angles of the spins change as well, with the fastest changes occurring in the vicinity of the phase transitions. Thus, the phase transitions at $\theta_c$ are not discontinuous, rather they appear to be a zero temperature, finite model analog of a second order phase transition, although the continuous transition occurs over a rather narrow range of $\theta$.

The complete phase diagram of the ground state of our 5-site Hubbard bowtie and other finite and extended triangular lattice systems
is of interest in and by itself, especially with respect to their symmetries.
A more complete formal analysis of the phase boundaries and other symmetry-related properties will be the subject of a forthcoming study.

\subsection{\label{sec:method3} Exchange-only SDFT}

Exact exchange in noncollinear SDFT has been defined in Ref. \cite{Ullrich2018}. Starting point is the exchange energy
\begin{equation}
 E_{\rm x} = - \frac{1}{2} \int \int \frac{d \bfr d\bfr'}{|\bfr-\bfr'|} \mathrm{Tr}\Big[\tens{\gamma}(\bfr,\bfr') \tens{\gamma}(\bfr',\bfr) \Big] .
 \label{eq:energy_exchange_hole}
\end{equation}
Here, $\tens{\gamma}$  denotes the one-particle spin-density matrix, a $2\times 2$ matrix in spin space whose elements are given by
$\gamma_{\sigma\xi}(\bfr,\bfr') = \sum_j^N \psi_{j\sigma}(\bfr)\psi^*_{j\xi}(\bfr')$,
constructed from two-component spinor Kohn-Sham orbitals, where $\sigma = \uparrow,\downarrow$ and likewise for $\xi$;
$\mathrm{Tr}$ is the trace over spin indices.
The exact noncollinear exchange potential then follows by minimizing $E_{\rm x}$ with respect to the orbitals, under the constraint that the orbitals
come from a single-particle equation with a local potential---this is the so-called optimized effective potential (OEP) approach \cite{Kummel2008}.
This approach is system-independent, i.e., it can be defined in real space and for lattice models alike.

The exact-exchange OEP requires solving an integral equation; we use here instead a simplification known as the Krieger-Li-Iafrate (KLI) approximation \cite{KLI92}.
The construction and numerical solution of the noncollinear KLI approximation have been discussed in detail in Refs. \cite{Ullrich2018,Tancogne2022}.
KLI directly yields a scalar exchange potential and an exchange magnetic field with
moderate numerical effort and with very little loss of accuracy compared to the full OEP.
In time-dependent SDFT, the exact-exchange OEP formally carries a memory \cite{Wijewardane2008}. The time-dependent KLI, on the other hand,
is an adiabatic approximation.

KLI for noncollinear systems produces exchange torques in extended Hubbard systems \cite{Pluhar2019}. For the purposes of the present
study, we also define a projected KLI (KLIp) in which the exchange magnetic field $\bfB_{\rm x}$  on each lattice site is projected
along the local magnetization direction, and which therefore has no exchange torques.

\subsection{SDFT phase diagram}

In the SDFT modeling of the Hamiltonian (\ref{eqn:ham}), a similar magnetic phase diagram is obtained as the exact one shown in Fig.~\ref{fig:PD}a.
The main difference is that the phase boundaries at the critical angles $\theta_c$ are not as sharp as in the exact case but quite diffuse, as schematically
depicted in Fig.~\ref{fig:PD}b. This is mainly due to the well-known tendency of SDFT to prefer symmetry breaking, unless highly accurate correlation functionals
are used.

The broadened phase boundary region has a tendency to exhibit ``charge sloshing'' \cite{Zhou2018} in the Kohn-Sham self-consistency iterations.
Charge sloshing spoils the convergence behavior and must be overcome with special measures, e.g. charge preconditioning or imaginary time propagation \cite{Flamant2019}.
A sufficiently strong external potential $V_j$ can also be applied to one side of the model in order to prevent charge sloshing. A fairly strong external potential in the exchange-only SDFT modeling is also necessary in the vicinity of $\theta_c$ in order to match to the exact initial state because correlation effects tend to be stronger close to the phase boundaries (see Sec. \ref{sec:III}).

For the simulations of section \ref{sec:D}, where the SDFT calculations are not tethered to an exact initial solution, charge sloshing can arise in the stronger interaction regime, even far from the critical angle $\theta_c$. We found that replacing the Kohn-Sham  self-consistency loop with an imaginary time propagation algorithm \cite{Flamant2019} for computing the SDFT ground state was useful in mitigating charge sloshing.

\section{Time propagation and choice of initial state} \label{sec:III}

In order to compare the dynamics of the exact and TD-SDFT solutions, we excite the system with a small, localized magnetic field burst along the $y$ direction during a brief number of time steps. To propagate the full time-dependent many-body Schr\"odinger equation for our
Hubbard bowtie we use a standard Crank-Nicolson algorithm. The time-dependent Kohn-Sham equations are also propagated using 
Crank-Nicolson, including a pre\-dic\-tor-corrector scheme (one corrector step suffices) \cite{Ullrich2012}.

Since our interest is predominantly in the dynamical effects comparing KLI and KLIp, we start in both cases from the same ground state.
This means that the exchange torques must be included in the calculation of the KLIp initial state, as this is required in order to have KLIp start with the same initial conditions as the full KLI simulations; however, these torques are frozen in, effectively in the form of an external magnetic field. By contrast, in full KLI the exchange torques are time-dependent as the system evolves.

Compared to the differences between exchange-only SDFT and exact many-body benchmarks, the differences between KLI and KLIp are small and
can easily be overshadowed. Since we are here interested in relatively subtle dynamical exchange torque effects, it is desirable to start from
a KLI initial state with external scalar potential $V_j$ and magnetic field $\bfB_j$
chosen to reproduce the exact density and magnetization. With some effort, $V_j$ and $\bfB_j$ can be numerically constructed by minimizing the functional
\begin{equation}
    F(V_j,\bfB_j ) = \sum_j \left[ (n_j-n_j^{(0)})^2 + |\bfm_j -\bfm_j^{(0)}|^2 \right],
\label{eqn:dist}
\end{equation}
where $n_j^{(0)}$ and $\bfm_j^{(0)}$ are the target density and magnetization, respectively. For each simulation matched to an exact initial state, we minimize $F$ to an accuracy of at least $F=10^{-25}$. The minimization is done via a conjugate gradient method with randomized resets when a local minimum of insufficient accuracy is reached. Searching over $V_j$ and $\bfB_j$ of only the SDFT simulations to find the minimum of $F$ is extremely computationally expensive due to the high dimensionality of the parameter space. In order to overcome this issue, we switch to minimizing $F$ with respect to the external fields of the exact solution once $F \lesssim 10^{-4}$. Minimizing with respect to exact solution parameters is less computationally expensive due to the much smoother response of the exact solution to small changes in the external fields.

\begin{table}
    \centering
    \caption{\label{table} SOC angle $\theta$ and interaction strength $U_0$ for the three ground states considered in Sec. \ref{sec:res4},
    the total magnitude of the exact xc torque and the exchange-only torque, and the correlation and exchange energies.}
    \begin{ruledtabular}
    \begin{tabular}{c c c c c c}
        $\theta$ & $U_{0}$ & $\Sigma_{j} |\bm{\tau}_{\rm xc}|$ & $\Sigma_{j} |\bm{\tau}_{\rm x}^{\rm KLI}| $ & $E_{\rm c}$ & $E_{\rm x}$ \\ \hline
        \rule{0mm}{4mm}
        $30^{\circ}$ &  1  & $4.2 \times 10^{-2}$ &  $2.6 \times 10^{-2}$    & -0.214  & -1.84  \\
        $30^{\circ}$ & 3 & $4.8 \times 10^{-2}$  & $1.2 \times 10^{-1}$  & -0.236 & -5.82 \\
        $60^{\circ}$ & 1 & $1.3 \times 10^{-4}$ & $2.0 \times 10^{-3}$  & -0.448 & -1.92  \\
    \end{tabular}
    \end{ruledtabular}
\end{table}

\section{\label{sec:res4} Results and discussion}

The model system shown in Fig. \ref{fig:geo} is simple yet exhibits quite a rich range of structural and dynamical behavior.
The parameter space to be explored comprises the hopping strength $t_h$, the SOC angle $\theta$, and the interaction strength $U_0$ 
(fixing $U_1 = U_0/2$). In the following we set $t_h=1$ and limit ourselves to three representative choices of $(\theta,U_0)$ in 
the magnetic phase diagram. This will already be sufficient to gain insight into the significance of the xc torques.

Table \ref{table} gives an overview of the three parameter sets, the ground-state exchange and correlation energies $E_{\rm x}$ and 
$E_{\rm c}$, and the magnitude of the exact xc torque $\bm{\tau}_{\rm xc}$ and of the exchange-only torque $\bm{\tau}_{\rm x}$. These will be further discussed below.

\subsection{\label{sec:A}$\theta =30^{\circ}$, $U_0 = 1$}

\begin{figure}
\includegraphics[width=\columnwidth]{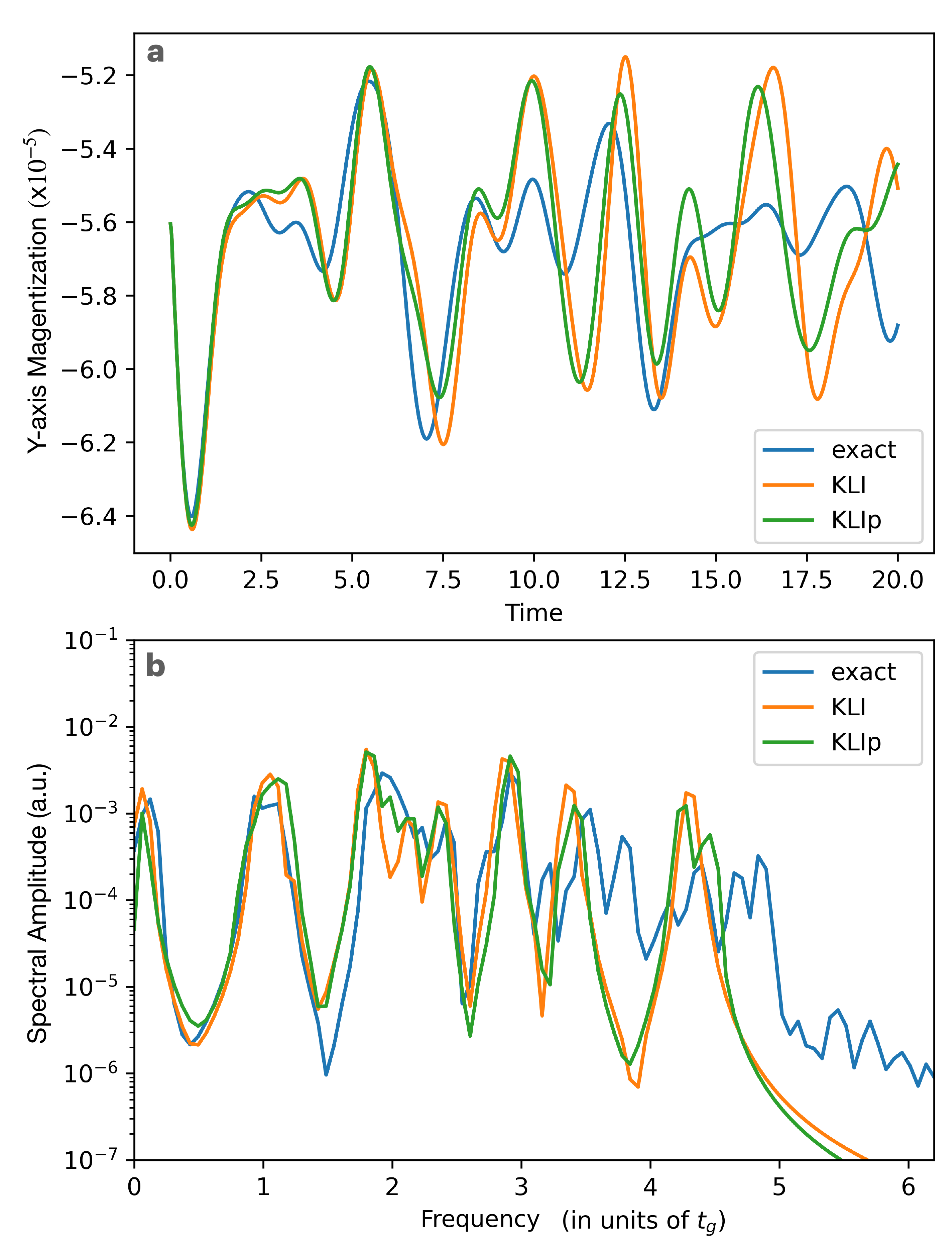}
\caption{\label{fig:A} Comparison of exact, KLI, and KLIp modeling for the case of $\theta =30^{\circ}$ and $U_0  = 1$. (a) Dynamics of the $y$-component of the
magnetization of a corner site exited by a small, short, local burst of magnetic field in the $y$ direction. (b) Associated spectral amplitude (in arbitrary units),
calculated via Fourier transform of the data shown in part (a).}
\end{figure}

We first consider the case $\theta =30^{\circ}$, which is in the middle of the spin-frustrated region shown in yellow in the phase diagrams
of Fig. \ref{fig:PD},
and for weak interaction strength $U_0  = 1$. The magnetization dynamics comparison of exact, KLI, and KLIp is shown in Fig.~\ref{fig:A}a, which
depicts the magnetization along the $y$-direction of a corner site. By construction (see Sec. \ref{sec:III}), all three methods start from the same initial value.

KLI and KLIp stay fairly close to one another for much of the run time due to the relative smallness of the Hubbard interaction, which indicates that the exchange torques
are not very important in the chosen regime. For the first few cycles of the precessional motion triggered by the short pulse, exchange-only SDFT is quite close
to the exact result. In spite of that, both KLI and KLIp start to diverge significantly from the exact solution around $t=15$, which shows that the correlation effects,
although relatively small, eventually start playing a nonnegligible role in the time evolution of the system.

To gain further insight, we perform a spectral analysis of the time-dependent data via Fourier transformation of the amplitude of the magnetization oscillations,
which reveals the spectrum of magnetic excitations. As shown in Fig.~\ref{fig:A}b, KLI and KLIp agree well with the exact spectrum at low frequencies
(up to about a frequency $\omega=3$). At higher frequencies, the SDFT spectra differ from the exact spectra, which may be due to the fact that we are
using here an adiabatic approximation which does not produce double or higher excitations \cite{Ullrich2012} and hence does not capture all peaks.
However, KLI and KLIp remain very close to each
other throughout, illustrating again that exchange torques are insignificant here.

\subsection{\label{sec:B}$\theta =30^{\circ}$, $U_0 = 3$}

\begin{figure}
\includegraphics[width=\columnwidth]{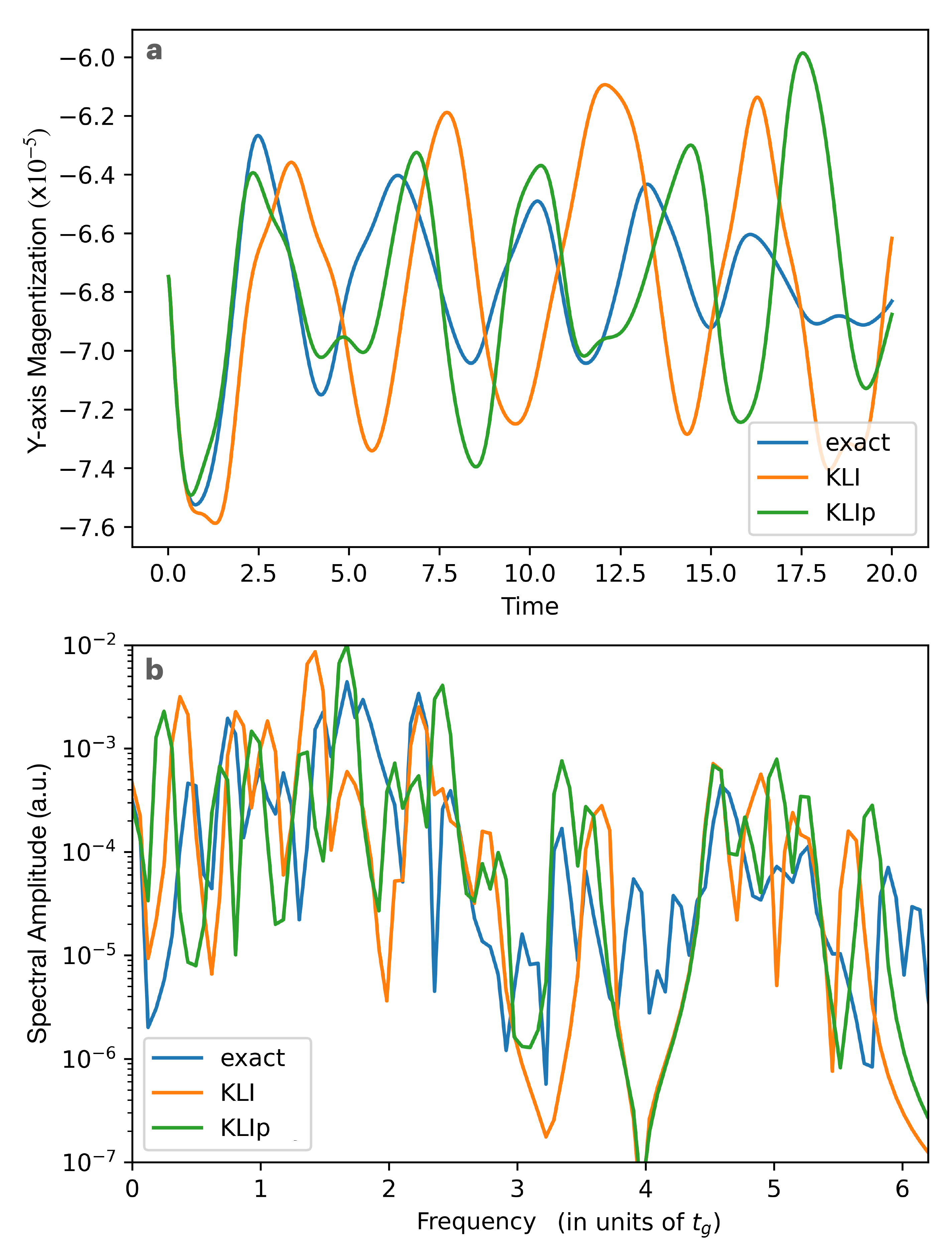}
\caption{\label{fig:B} Same as Fig. \ref{fig:A} but for $U_0  = 3$.}
\end{figure}

For the second case, we remain at $\theta =30^{\circ}$, away from the phase boundaries, but increase the interaction strength into the moderately strongly
interacting regime, at $U_0 = 3$. The real time magnetization dynamics and amplitude spectrum are shown in Fig.~\ref{fig:B}.
Clearly, KLI and KLIp start to differ from each other almost right away, which points to the more important role of the exchange torques.

At first glance, it is surprising to see that the projected KLI, which has no torques, agrees better with the exact magnetization oscillations, at least
for the first few cycles.
To explain this, it is helpful to consider the magnitudes of the initial $\bm{\tau}_{\rm xc}$ and $\bm{\tau}_{\rm x}$ given in Table \ref{table}.
For $U_0=1$, the sum of the exchange torques is comparable to the sum of the xc torques (within a factor 1.6); at $U_0=3$, on the other
hand, the exchange torques are much larger than the xc torques, which suggests that the correlation contribution to the torques becomes relatively
much more important. In other words, exchange-only overestimates the torques, and correlation compensates for it. KLIp avoids this
overestimation (better no exchange torque at all, than too much of it), and brings the dynamics closer to the exact case.
Notice that this could have not been anticipated just from looking at the exchange and correlation energies $E_{\rm x}$ and $E_{\rm c}$ of
the initial state, which would have suggested that the exchange is dominant.

The Fourier spectrum in Fig.~\ref{fig:B}b is less clear: while both KLI and KLIp seem to reproduce the rough trends of the exact spectrum,
it is difficult to say which one of them agrees better. Neither of them captures the details of the exact spectrum particularly well.

\subsection{\label{sec:C}$\theta =60^{\circ}$, $U_0 = 1$}

\begin{figure}
\includegraphics[width=\columnwidth]{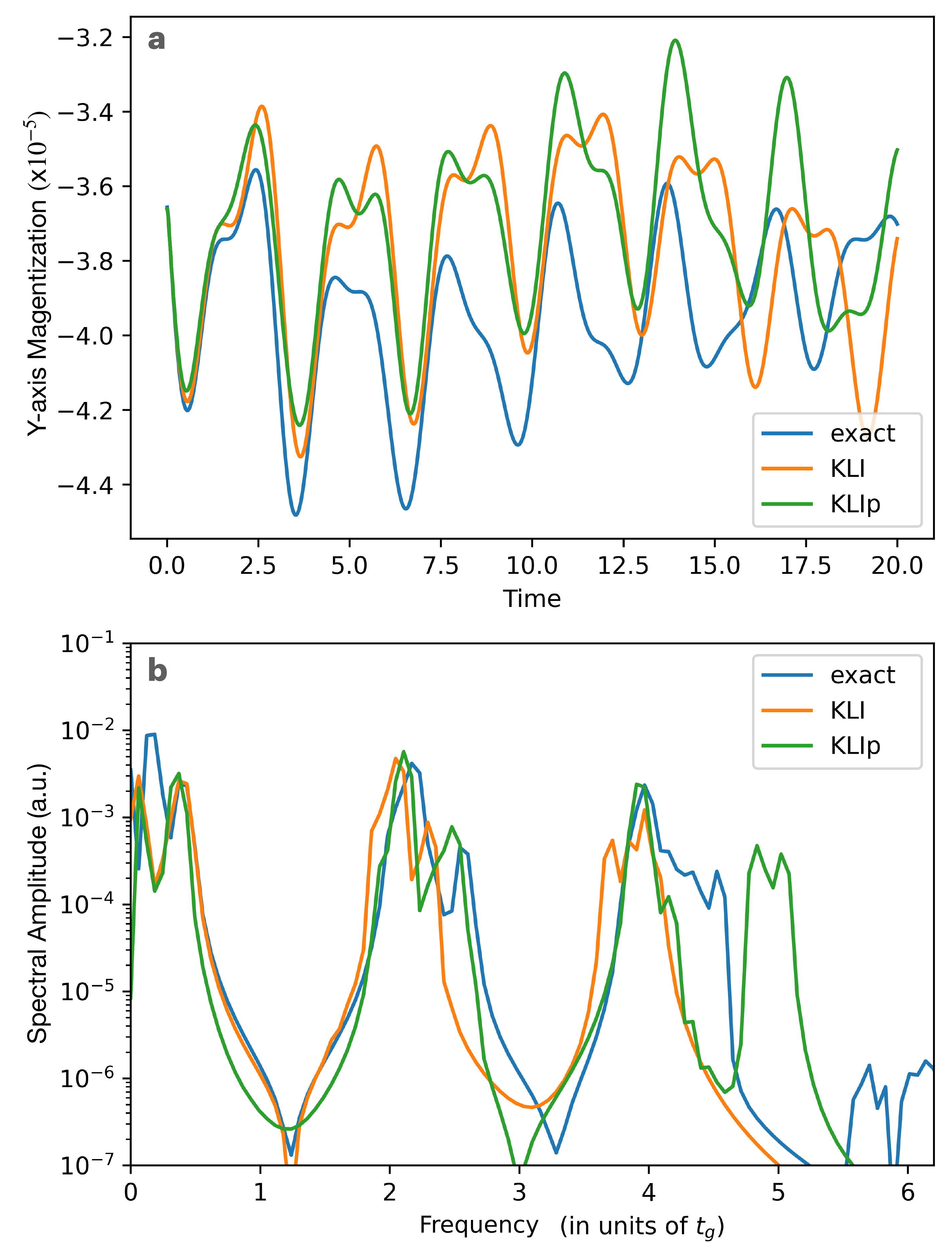}
\caption{\label{fig:C} Same as Fig. \ref{fig:A} but for $\theta =60^{\circ}$. }
\end{figure}

Lastly, we consider the case of $\theta =60^{\circ}$ and $U_0  = 1$, see Fig.~\ref{fig:C}. This state is at a critical angle of the magnetic phase diagram where artificial charge density symmetry breaking in exchange-only SDFT is prevalent, indicating that strong correlations are
needed to reproduce the exact results. As shown in Table \ref{table}, $E_{\rm c}$ is significantly enhanced relative to $E_{\rm x}$,
compared to the case of $\theta=30^\circ$. Correspondingly,
the exchange torques are lower, due to the localization of the magnetization to one side of the system. The strong correlation effects at the transition angle result in both KLI and KLIp diverging from the exact solution fairly quickly. The magnetization oscillations calculated with KLI and KLIp match each other fairly well, at least for the first few cycles, but then differences start to accumulate.

The Fourier spectrum, see Fig.~\ref{fig:C}b, has well defined excitations, which are fairly well captured by both KLI and KLIp,
but some inaccuracies are noticeable at both high and low frequencies. Notably, KLIp performs slightly better at estimating the gaps in the spectrum for mid-range frequency excitations. The better performance of KLIp occurs, similarly to Section \ref{sec:B}, due to the KLI
exchange-only approximation substantially overestimating the xc torques, with no correlation to compensate (see Table \ref{table}).

\begin{figure}
\includegraphics[width=\columnwidth]{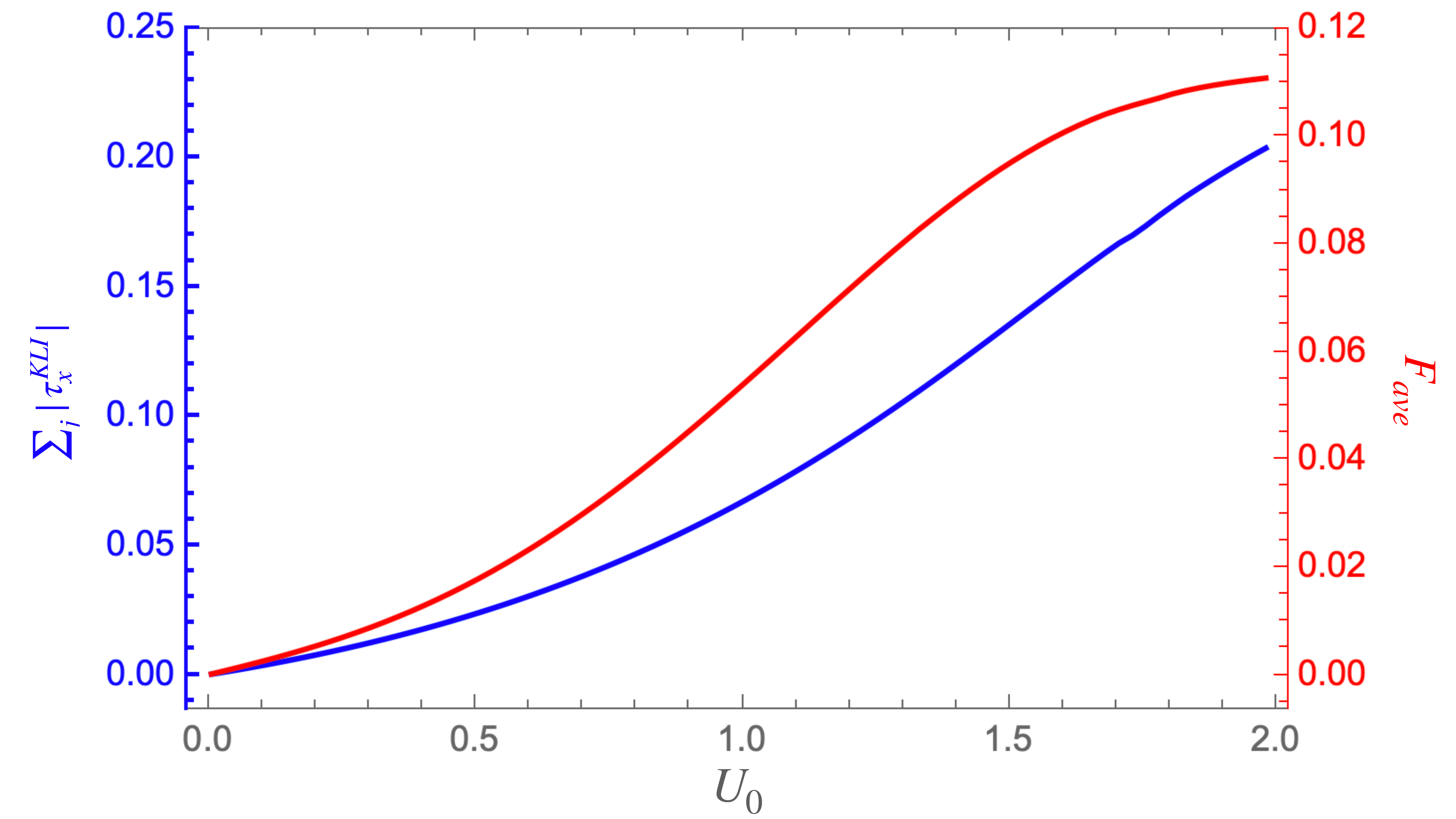}
\caption{\label{fig:D} Red (right axis): Comparison between KLI and KLIp solutions as a function of interaction strength for the case of $\theta =30^{\circ}$ and $U_0$, quantified by the time-averaged distance measure $F_{\rm ave}$, Eq. (\ref{F_ave}).
Blue (left axis): $\Sigma_{j} |\bm{\tau}_{\rm x}^{\rm KLI}| $ of the Hubbard bowtie ground state versus $U_0$. 
}
\end{figure}

\subsection{\label{sec:D} Distance between KLI and KLIp versus $U_0$}

The effect of the exchange torques can be further quantified by introducing the time-averaged distance measure
\begin{eqnarray}\label{F_ave}
F_{\rm ave} &=& \frac{1}{t}\int_0^t dt' \sum_j
 \bigg[ \left(n_j^{\rm KLI}-n_j^{\rm KLIp}\right)^2 \nonumber\\
 &+& \left|\bfm_j^{\rm KLI} -\bfm_j^{\rm KLIp}\right|^2 \bigg]\:,
\end{eqnarray}
where we calculate the time average over a short time ($t=2$) after initial excitation. This provides an estimate of the degree of divergence between the solutions which can be compared with interaction strength and the magnitude of ground state KLI exchange torques.

Figure \ref{fig:D} shows the time-averaged distance measure (\ref{F_ave}) between KLI and KLIp as a function of $U_0$ at $\theta =30^{\circ}$, and,
for the sake of comparison, the sum of the magnitudes of the KLI exchange torques of the corresponding initial states.
Both $F_{\rm ave}$ and $\Sigma_{j} |\bm{\tau}_{\rm x}^{\rm KLI}| $ start out linearly for small interaction strengths
$U_0$ and keep increasing well into the moderate interaction regime, where $F_{\rm ave}$ appears to start leveling off around $U_0=2$.

A comparison with exact time-dependent xc torques is, unfortunately, not possible; even  the construction of the exact
$\Sigma_{j} |\bm{\tau}_{\rm xc}|$ over the whole range of $U_0$ is numerically too demanding, except for the three cases
in Table \ref{table}. Nevertheless, we can infer from the results presented in Fig. \ref{fig:D} that both exchange and correlation torques must be accounted for even for relatively low interaction strengths in order to accurately describe the dynamics.

\section{\label{sec:con5}conclusion}

We have performed exact and approximate, exchange-only (TD)-SDFT calculations on a half-filled 5-site Hubbard cluster
with varying interaction and SOC strengths. The purpose of this study was to assess the significance of many-body magnetic torques for the
description of spin dynamics. We considered three scenarios with weak and moderate interactions and close to and away from
a transition between different magnetic phases. While this is clearly not an exhaustive exploration of the parameter space, the examples studied
here are good representatives and allow us to draw meaningful conclusions.

We find that exchange torques become increasingly important as non-local interactions become stronger, with an approximately linear dependence at low interactions (see Fig.~\ref{fig:D}), but the relationship becomes nonlinear for more general interaction strengths. Strong correlations in the vicinity of phase boundaries reduce the importance of exchange torques due to localization. When correlations are particularly strong, they appear to counteract the exchange torques, leading to a net reduction of the total xc torques.
This suggests that when lacking a sufficiently accurate correlation functional, completely projecting out the xc torques may improve the overall
accuracy of TD-SDFT magnetic dynamics, at least for short times.

The challenge for future work is clearly to construct correlation functionals that produce accurate torques, and test these against benchmarks.
A good starting point will be to do this for similar finite Hubbard models, followed by tests for the magnetization dynamics
in real magnetic materials in the linear and nonlinear regime.

\acknowledgments

This work was supported by DOE Grant No. DE-SC0019109. The authors wish to thank Aurora Pribram-Jones for helpful discussion.

\bibliography{noncol_ref}

\end{document}